\documentclass[
    prx,
    twocolumn,
    longbibliography
]{revtex4-1}

\usepackage{float}
\usepackage{bbm}
\usepackage{epsfig}
\usepackage{epstopdf}
\usepackage{graphicx, subfigure}
\usepackage{amssymb}
\usepackage{amsmath}
\usepackage{color}
\usepackage{hyperref}
\usepackage[capitalize]{cleveref}
\usepackage{braket,mleftright}
\usepackage{physics}
\usepackage{import}
\usepackage{dcolumn}
\usepackage{csquotes}

\floatstyle{plaintop}
\restylefloat{table}

\begin{document}

    \title{Microwave-based Arbitrary {\sc cphase} Gates for Transmon Qubits}

    \author{George S. Barron$^1$}
    \author{F. A. Calderon-Vargas$^1$}
    \author{Junling Long$^2$}
    \author{David Pappas$^2$}
    \author{Sophia E. Economou$^1$}

    \affiliation{$^1$Department of Physics, Virginia Tech, Blacksburg, Virginia 24061, USA \\ $^2$National Institute of Standards and Technology, 325 Broadway, Boulder, CO 80305--3328, USA}

    \begin{abstract}
        Superconducting transmon qubits are of great interest for quantum computing and quantum simulation.
        A key component of quantum chemistry simulation algorithms is breaking up the evolution into small steps, which naturally leads to the need for non-maximally entangling, arbitrary {\sc cphase} gates.
        Here we design such microwave-based gates using an analytically solvable approach leading to smooth, simple pulses.
        We use the local invariants of the evolution operator in $SU(4)$ to develop a method of constructing pulse protocols, which allows for the continuous tuning of the phase.
        We find {\sc cphase} fidelities of more than $0.999$ and gate times as low as $100\text{ ns}$.
    \end{abstract}

    \maketitle

    \section{Introduction}\label{sec:intro}

    Quantum computing promises solutions to a number of problems in computing, chemistry, and material science.
    Superconducting qubits are a promising candidate for qubits because their fabrication relies on existing techniques~\cite{frunzio_fabrication_2005,majer_coupling_2007}, and they can also have their characteristics tailored for specific applications.

\begin{figure}[h]
        \centering
        \includegraphics[width=0.6 \columnwidth]{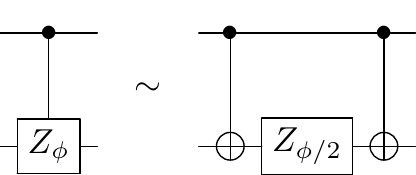}
        \caption{Non-maximally entangling {\sc cphase} gate decomposed into two maximally entangling {\sc cnot} gates.}
        \label{fig:circ}
    \end{figure}

    Superconducting qubits have been recently used in the implementation of quantum algorithms for molecular problems~\cite{kandala_hardware-efficient_2017, colless_computation_2018,omalley_scalable_2016}, reinforcing the idea that quantum chemistry is one of the most appealing applications of quantum computing~\cite{preskill_quantum_2018}.
    In many quantum simulation algorithms, gate decompositions of Trotterized Hamiltonians often include {\sc cphase} gates, which are then written in terms of two maximally entangling {\sc cnot} gates~\cite{georgescu_quantum_2014}.
    This decomposition is shown in Fig.~\ref{fig:circ}.
    Clearly, using {\sc cphase} gates instead of {\sc cnot}s would reduce circuit depth and potentially improve resource use in terms of time and fidelity.

    Fast high-fidelity two-qubit gates remain challenging in superconducting qubits~\cite{gu_microwave_2017}.
    Spectral crowding makes accurately addressing an individual transition to produce a controlled operation difficult over short times because the bandwidth required to resolve differences between nearby transitions becomes very small, increasing the time required for each gate~\cite{theis_simultaneous_2016}.
    The trade-off is then that either gate times are long or the gate fidelity is low.

    One approach to implementing two-qubit gates in superconducting qubits is to dynamically tune elements of the circuit.
    For example, one can either tune the qubit frequency~\cite{majer_coupling_2007,dicarlo_demonstration_2009,dicarlo_preparation_2010,ghosh_high-fidelity_2013,martinis_fast_2014,barends_superconducting_2014}, resonator frequency~\cite{mckay_universal_2016,roth_analysis_2017}, or the coupling strength~\cite{chen_qubit_2014,royer_fast_2017}.
    Unfortunately, tunable elements introduce charge noise, leading to decoherence and low fidelity.
    An alternative method is to apply microwave pulses to the qubits to drive transitions that implement unitary rotations~\cite{galiautdinov_generation_2007,chow_simple_2011, chow_microwave-activated_2013,economou_analytical_2015,sheldon_procedure_2016,paik_experimental_2016,deng_robustness_2017,allen_optimal_2017,barnes_fast_2017, noh_construction_2018,magesan_effective_2018,egger_entanglement_2019,tripathi_operation_2019,allen_minimal_2019,premaratne_implementation_2019}.
    Typically, microwave-based control selects a single transition to implement a two-qubit gate.
    However, spectral crowding is a generic issue for systems controlled exclusively by microwave pulses since, without tuning, their spectra are fixed (up to Stark shift effects) and this usually forces the gate time to be very long to spectrally select the target transition.
    Moreover, the always-on coupling in these systems makes single-qubit gates nontrivial, especially for strongly coupled qubits.

    In this work, we develop a collection of microwave-based {\sc cphase} gates using the SWIPHT (Speeding up Wave forms by Inducing Phases to Harmful Transitions)~\cite{economou_analytical_2015} protocol, which overcome spectral crowding.
    This protocol was recently used in experiment to produce {\sc cnot} gates between two transmon qubits~\cite{premaratne_implementation_2019}.
    Here, we make use of hyperbolic secant (sech) pulse envelopes~\cite{rosen_double_1932} which are smooth, simple to implement, and produce high fidelities with low gate times for a variety of angles~\cite{economou_proposal_2006,economou_high-fidelity_2012}.
    These type of pulses were recently used on transmons in experimental demonstrations of $Z$ gates~\cite{ku_single_2017} and as part of a two-qubit gate~\cite{noh_construction_2018}.
    We use the local invariants~\cite{makhlin_0000,Zhang2003} of the two-qubit analytic evolution operator with control sech pulses to find conditions on the pulse parameters that achieve the desired two-qubit operation.
    Through simulations of transmons with typical parameters, we show that our {\sc cphase} gates produce high fidelities for low gate times.
    These {\sc cphase} gates are applicable in either an all-microwave context or a microwave-tuning hybrid context.
    Regarding the latter, our {\sc cphase} gates are applicable in the sense that they only rely on a weak effective $Z\otimes Z$ coupling compared to methods that dynamically tune circuit elements.
    This reliance on only a weak amount of dynamical tuning of the circuit parameters allows these gates to be performed in a variety of parameter regimes.
    To address the generic challenge of implementing single-qubit gates with fixed-frequency, always-coupled transmons, we design a composite pulse protocol that gives high-fidelity $X$ rotations, which along with our two-qubit gates and previously available $Z$ gates~\cite{ku_single_2017,mckay_efficient_2017} form a universal set.
    These single-qubit gates all take less than $50\text{ ns}$ each and have fidelities in excess of $0.992$.

    This paper is organized as follows.
    In Section~\ref{sec:transmon-system} we introduce the two-qubit Hamiltonian for the system of transmons coupled by a resonator.
    In Section~\ref{sec:cphase} we present the results of the analytical {\sc cphase} protocols and numerical performance, as well as their robustness in other coupling strength regimes.
    In Section~\ref{sec:sq-gates} we present our single-qubit gates along with their fidelities.
    We conclude in Section~\ref{sec:conclusions}.

    \section{Transmon Hamiltonian}\label{sec:transmon-system}

    We focus on two superconducting transmon qubits coupled by a cavity~\cite{koch_charge-insensitive_2007}.
    The transmons are modeled as weakly anharmonic oscillators, and the cavity as a harmonic oscillator.
    The Hamiltonian for this system is given by
    \begin{align}
        H_0
        = &  \omega_c a^\dagger a + \sum_{j = 1,2} \epsilon_{j,1} a_j^\dagger a_j - \frac{\eta_j}{2} a_j^\dagger a_j ( a_j^\dagger a_j - 1 ) + \\
        & g_{j} \left( a_j^\dagger a + a^\dagger a_j \right). \nonumber
    \end{align}
    Here $\omega_c$ is the frequency of the cavity connecting the two qubits, $\epsilon_{j,1}$ is the transition frequency between the ground and first excited state for the         $j^\text{th}$ qubit, $\eta_j$ is the anharmonicity of the $j^\text{th}$ qubit, $g_j$ is the coupling strength between the cavity and the $j^\text{th}$ qubit, $a$     ($a^\dagger$)     is the annihilation (creation) operator for the cavity, and $a_j$ ($a_j^\dagger$) is the annihilation (creation) operator for the $j^\text{th}$ qubit.
    The Hamiltonian describing the coupling to the external microwave electric field is given by
    \begin{equation}
        H_p(t) = \sum_{j=1,2} E_j(t) e^{i\omega_{p,j}t} a_j + \text{H.c.},
    \end{equation}
    where $E_j(t)$ and $\omega_{p,j}$ are the pulse envelope and frequency driving the     $j^\text{th}$ qubit, respectively.
    For the design of our gates we only drive (without loss of generality) the second qubit so that $E_1(t) = 0$, $E_2(t) = E(t)$ and $\omega_{p,2} = \omega_p$.

    The states in the system are $\ket{i;j,k}=\ket{i}_c \ket{j,k}$ , where $\ket{i}_c$ is the     $i^\text{th}$ cavity level and the $j^\text{th}$ ($k^\text{th}$) index denotes the level of the first (second) transmon.
    It is advantageous to write out the Hamiltonian in the dressed basis~\cite{cohen-tannoudji_atom-photon_1992}, which diagonalizes     $H_0$, and the indices of each element of the dressed basis is determined by the state in the bare basis that has the largest overlap with the dressed state.
    For example, for indices $s_i$ we write an element of the dressed basis as an eigenstate of $H_0$ with     $\ket{\widetilde{s}_1} = \sum_{i} \alpha_i     \ket{s_i}$ where $|\alpha_1| > |\alpha_i|$ with     $i \neq 1$.
    We encode each qubit into the lowest two levels of each transmon.
    Consequently, the projection operator for the two-qubit subspace is     $P_\text{QSS} =     \ket{\widetilde{0;0,0}}\bra{\widetilde{0;0,0}} + \ket{\widetilde{0;0,1}}\bra{\widetilde{0;0, 1}} +     \ket{\widetilde{0;1,0}}\bra{\widetilde{0;1,0}} +     \ket{\widetilde{0;1,1}}\bra{\widetilde{0;1,1}}$.
    Going to the dressed basis and projecting into the qubit subspace spanned by the basis $\ket{\widetilde{0;0,0}}$,     $\ket{\widetilde{0;0,1}}$, $\ket{\widetilde{0;1,0}}$, $\ket{\widetilde{0;1,1}}$, the approximate two-qubit Hamiltonian when only one qubit is driven is given by
    \begin{widetext}
        \begin{equation}
            \label{eq:hqss}
            H_\text{QSS}  \approx \\
            \begin{pmatrix}
                -\omega_{I,1} / 2 &  \Omega_1(t) e^{i \omega_p t} & 0 & 0  \\
                \Omega_1(t)^* e^{-i \omega_p t} & +\omega_{I,1} / 2 & 0 & 0  \\
                0 & 0 & -\omega_{I,2} / 2 &  \Omega_2(t) e^{i \omega_p t} \\
                0 & 0 &  \Omega_2(t)^* e^{-i \omega_p t} & +\omega_{I,2} / 2
            \end{pmatrix}.
        \end{equation}
    \end{widetext}
    We define $\delta\omega_I$ as the difference between the transition frequencies of the two subspaces $\omega_{I,1}$ and $\omega_{I,2}$ each corresponding with subspace $1$     (upper left block) and subspace $2$ (lower right block) of the Hamiltonian, respectively, as well as $\Omega_i(t) = E(t) d_i$ for the dipole moment $d_i$ of each transition.
    Here we have made the approximation that terms in the Hamiltonian that couple states with a different number of excitations on the first qubit will vanish.
    This is due to the fact that in the dressed basis, since our off-diagonal coupling terms in $H_0$ are small compared to the diagonal terms, $\left|\braket{\widetilde{i;j,k}}{i;j, k}\right|$ is large compared to contributions from other states.

    To design fast gates, we avoid spectrally selecting one of the two subspaces and allow the pulse to drive both transitions.
    Because in general $d_1 \neq d_2$ and $\omega_{I,1}     \neq \omega_{I,2}$, the same $E(t)$ on each block will produce different evolutions.
    Our goal is to design control pulses $E(t)$ that generate two-qubit gates of the form     $\ket{0}\bra{0} \otimes \mathbb{I}_2 + \ket{1}\bra{1} \otimes U$, and other control pulses that generate single-qubit gates of the form $\mathbb{I}_2 \otimes U$.

    \section{{\sc cphase} Gates}\label{sec:cphase}
\begin{figure}
        \centering
        \includegraphics[width=\columnwidth]{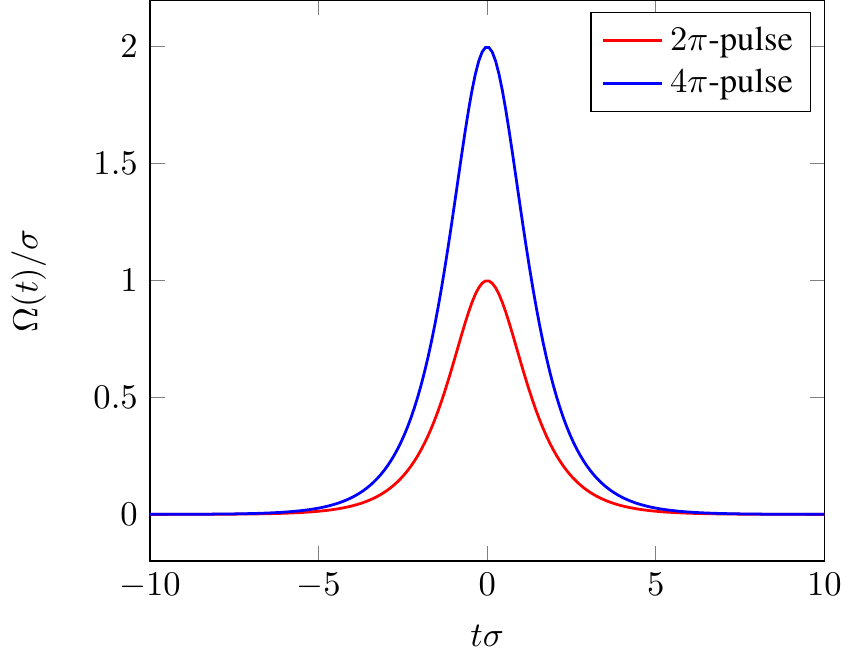}
        \caption{Hyperbolic secant $2\pi$ and $4\pi$-pulses.
        These two different pulse areas have different algebraic properties, which lead to different types of protocols.}
        \label{fig:sech}
    \end{figure}

    For each of the following {\sc cphase} gates, we use hyperbolic secant pulses of the form $\Omega(t) = \Omega_0 \sech(\sigma t)$ with bandwidth $\sigma$, amplitude $\Omega_0$, and pulse frequency $\omega_p$.
    This pulse is chosen because it gives an analytically solvable time-dependent Schr\"odinger equation for a two-level system~\cite{rosen_double_1932}, is smooth and has nice analytic properties for rotations about the $Z$ axis~\cite{economou_proposal_2006} (see Appendix~\ref{sec:app-sech} for the derivation of the evolution operator and discussion of its properties).
    Specifically, for detuning $\Delta$ and bandwidth $\sigma$, a $2\pi$ hyperbolic secant pulse will induce a phase $2 \arctan(\sigma/\Delta)$ and a $4\pi$ pulse will induce a phase $2 \arctan\left(     \frac{4\Delta/\sigma}{(\Delta/\sigma)^2-3} \right)$~\cite{economou_high-fidelity_2012}.
    A plot of two examples of hyperbolic secant pulses is shown in Fig.~\ref{fig:sech}.
    The main idea is that the same sech pulse acts on both (target and harmful) transitions, causing a cyclic evolution to each subspace.
    This assumes that the dipoles of the two transitions are the same, which is not strictly the case.
    Nevertheless, approximately equal dipoles, as is the case for the parameters here, suffice for high fidelities.
    Due to the different detunings of the two transitions from the pulse, each acquires a different phase.
    The choice of phases for the two transitions, which we can control through the bandwidth and frequency of the pulse, determines the specific {\sc cphase} gate.
    Since we focus on {\sc cphase} gates, we use $2\pi$ and $4\pi$-pulses, which only implement cyclic transitions between energy levels.
    Our pulses generate generalized {\sc cphase} gates, defined as $\textsc{cphase}'=\mathrm{diag}(e^{i\phi_{00}},e^{i\phi_{01}},e^{i\phi_{10}},e^{i\phi_{11}})$, which is equivalent to a regular {\sc cphase} gate, $\textsc{cphase}=(1,1,1,e^{i\theta})$, up to local $Z$ rotations.
    The phases in both the generalized and regular {\sc cphase} gates satisfy     $\theta=\phi_{00}-\phi_{01}-\phi_{10}+\phi_{11}$.
    In systems of transmons, it has been shown that zero-duration single-qubit $Z$ rotations may be accomplished by shifting the phase of the microwave pulse~\cite{mckay_efficient_2017}, so this generalization does not affect our gate times or fidelities.
    Moreover, as discussed in Appendix~\ref{sec:app-sech}, the pulse areas considered here produce no transfer of population and hence only perform rotations about the $Z$ axis.
    For this reason, although the local invariants allow us to consider arbitrary evolutions in $SU(4)$ up to arbitrary rotations in $SU(2)$ (see Appendix~\ref{sec:mi}), our pulses only require that we consider local operations of the form $R_Z(\phi)$, which, as discussed above, do not affect gate times or fidelities.

\begin{figure}
        \centering
        \includegraphics[width=\columnwidth]{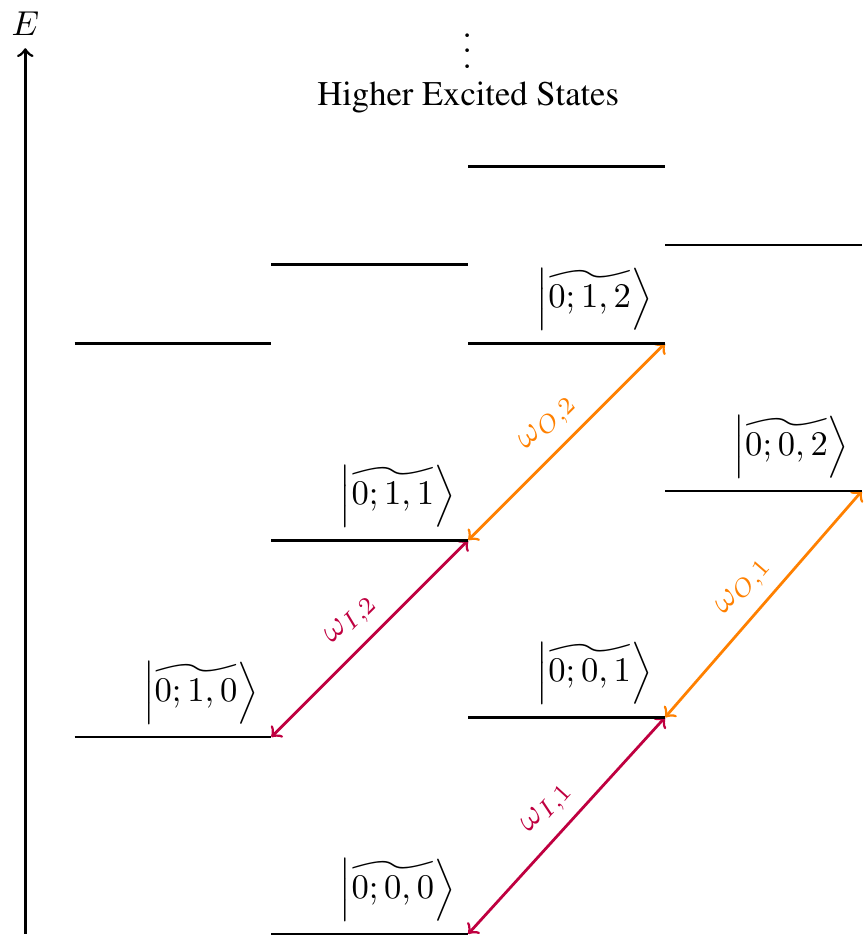}
        \caption{Transitions between two-qubit states.
        The purple transitions correspond to the IQSS protocols, and the orange transitions correspond to the OQSS protocols.}
        \label{fig:elevels}
    \end{figure}

    In the following results, we denote protocols that use transitions that exist inside the qubit subspace as ``IQSS", and protocols that use transitions partially outside the qubit subspace as ``OQSS". These two sets of transitions are illustrated in Fig.~\ref{fig:elevels}.
    In particular, when we refer to a protocol that is ``IQSS", the transitions and their respective frequencies that we consider are $\omega_{I,1} : \ket{\widetilde{0;00}} \leftrightarrow \ket{\widetilde{0;01}}$, $\omega_{I,2} :     \ket{\widetilde{0;10}} \leftrightarrow \ket{\widetilde{0;11}}$.
    On the other hand, if the protocol is ``OQSS", then the transitions and their respective frequencies that we consider are $\omega_{O,1} : \ket{\widetilde{0;01}} \leftrightarrow \ket{\widetilde{0;02}}$, $\omega_{O,2} : \ket{\widetilde{0;11}} \leftrightarrow \ket{\widetilde{0;12}}$.
    As per the SWIPHT protocol, in either of these cases we designate either the IQSS or OQSS transitions with either the harmful or target transitions with transition frequencies $\omega_{x,h}$ and $\omega_{x,t}$, respectively.
    From these we define the difference $\delta\omega_x = \omega_{x,t} - \omega_{x,h}$ with $x \in \left\{ I,O \right\}$ depending on the transitions chosen.

    When evaluating the performance of the derived protocols, we numerically solve the Schr\"odinger equation to obtain the evolution operator at the end of each pulse.
    In our simulations we keep $3$ states for the cavity and $4$ states for each of the qubits, so that the Hilbert space simulated is $48$-dimensional.
    This sufficiently simulates the full dynamics of the system in that adding more available states does not change our resulting fidelities.
    To compare the final evolution operator we obtain from the simulation with the target one, we calculate the fidelity given by ${F = \frac{1}{n(n+1)} \left( \text{Tr}\left(M M^\dagger\right) + \left| \text{Tr}\left(M\right) \right|^2 \right)}$~\cite{pedersen_fidelity_2007} where $M = U_0^\dagger U$, with $U_0$ being the desired gate and $U$ being the actual gate from simulations.
    Each $U$ and $U_0$ are truncated so that they act only on the qubit subspace.
    In our numerical simulations we use
    $\omega_c = 7.15 \text{ GHz}$,
    $\epsilon_{1,1} = 6.2 \text{ GHz}$,
    $\epsilon_{2,1} = 6.8 \text{ GHz}$,
    $\eta_1 = \eta_2 = \eta = 350 \text{ MHz}$,
    $g_1 = g_2 = g = 130 \text{ MHz}$ as the fixed parameters, except in Section~\ref{subsec:gate-via-resonant} where we evaluate the performance of the gates when varying the coupling strength $g$.
    From these, we find that $\delta\omega_I = 3.23 \text{MHz}$ and $\delta\omega_O = -11.07 \text{MHz}$.
    In our simulations, we truncate the sech pulses by switching the pulse on for time $10 / \sigma$.
    Moreover, we numerically optimize around the analytically predicted solution to compensate for errors such as a difference in the dipoles of the two transitions.
    Below we describe each of the protocols, and provide results from numerical simulations quantifying their performance.

    \subsection{{\sc cphase} gate via off-resonant $2\pi$-pulse OQSS}\label{subsec:gate-via-off-resonant-pulse-oqss}

    Our strategy here is to find conditions on the bandwidth and pulse frequency of a hyperbolic secant pulse that performs a generalized {\sc cphase} gate on the two subspaces defined above.
    To this end, we use the local invariants (Appendix~\ref{sec:mi}) of the analytical evolution operator for the two-qubit system driven by sech pulses (Appendix~\ref{sec:app-sech}).
    
    First, we recall that the definition of the generalized {\sc cphase} gate is $\textsc{cphase}' = \text{diag}\left( e^{i \phi_{00}} , e^{i     \phi_{01}} , e^{i \phi_{10}} , e^{i \phi_{11}} \right)$ where the phase imparted is $\theta = \phi_{00} - \phi_{01} - \phi_{10} + \phi_{11}$.
    If we consider the two block-diagonal portions of the Hamiltonian $H_{QSS}$~\eqref{eq:hqss}, we can define two detunings between the pulse and the desired transitions, $\Delta_1 =  \omega_p - \omega_{O,1}$ and $\Delta_2 =  \omega_p - \omega_{O,2}$.
    
    Now, for the OQSS protocol the block-diagonal form of the Hamiltonian as well as the analytical solution for the unitary operator of a two-level system discussed in Appendix~\ref{sec:app-sech} allow us to write the evolution operator as $U = \text{diag}\left( 1 , f_1 , 1, f_2 \right)$.
    Here ${f_{j} = {_2}F_1\left(-a,+a,\frac{-i \Delta_j + \sigma}{2\sigma}, 1\right)}$ is the Gaussian hypergeometric function with $a\equiv \frac{\Omega}{\sigma}\in \mathbb{Z}$.
    Using Eq.~\eqref{eq:local_invariants} We can compute the local invariants of the two-qubit evolution operator $U$, yielding
    \begin{equation}\label{eq:loc_inv_U_off_resonant}
    \begin{aligned}
        G_1(U) = & \frac{1}{8}(4+\gamma+\gamma^*), \\
        G_2(U) = & 0, \\
        G_3(U)= & 2+\frac{\gamma}{2},
        \end{aligned}
    \end{equation}
    where
    \begin{equation}
    \begin{aligned}
        \gamma =&\frac{\Gamma^2 \left(\frac{\sigma -i \Delta _1}{2 \sigma }\right) \Gamma \left(-a+\frac{1}{2}-\frac{i \Delta _2}{2 \sigma }\right) \Gamma \left(a+\frac{1}{2}-\frac{i \Delta _2}{2 \sigma }\right)}{\Gamma^2 \left(\frac{\sigma -i \Delta _2}{2 \sigma }\right) \Gamma \left(-a+\frac{1}{2}-\frac{i \Delta _1}{2 \sigma }\right) \Gamma \left(a+\frac{1}{2}-\frac{i \Delta _1}{2 \sigma }\right)}\\
        &+\frac{\Gamma^2 \left(\frac{\sigma -i \Delta _2}{2 \sigma }\right) \Gamma \left(-a+\frac{1}{2}-\frac{i \Delta _1}{2 \sigma }\right) \Gamma \left(a+\frac{1}{2}-\frac{i \Delta _1}{2 \sigma }\right)}{\Gamma^2 \left(\frac{\sigma -i \Delta _1}{2 \sigma }\right) \Gamma \left(-a+\frac{1}{2}-\frac{i \Delta _2}{2 \sigma }\right) \Gamma \left(a+\frac{1}{2}-\frac{i \Delta _2}{2 \sigma }\right)}.
        \end{aligned}
    \end{equation}
     We also compute the local invariants for the target {\sc cphase} gate, yielding
        \begin{equation}\label{eq:local_inv_cphase}
            \mathbf{G}(\text{diag}\left( 1,1,1, e^{i\theta} \right)) = \left( \cos(\theta/2)^2 , 0 , 2 + \cos(\theta) \right),
        \end{equation}
    where $\mathbf{G}(U)=\left( G_1(U),G_2(U),G_3(U)\right)$.
    
    In order to find the conditions on the pulse parameters, we demand that the local invariants of $U$ (Eq.~\eqref{eq:loc_inv_U_off_resonant}) be equal to those of the {\sc cphase} gate, and thus we arrive at (assuming for simplicity that $a=1$)
    \begin{equation}\label{eq:angle_condition_OQSS_off_resonant}
    \cos\theta = \cos\left( \phi_1(\Delta_1) - \phi_1(\Delta_2) \right),
    \end{equation}
    where $e^{i \phi_1(\Delta_j)} = \frac{\sigma - i \Delta_j}{\sigma + i \Delta_j}$.
    
    For the IQSS protocol, we follow the same procedure, except that in that case the IQSS evolution operator is ${U = \text{diag}\left( f_1 , f_1^* , f_2, f_2^* \right)}$
    After imposing that the respective local invariants of $U$ and the {\sc cphase} gate be equal, for a $2\pi$-pulse we arrive at 
    \begin{equation}
    \cos\theta = \cos\left( 2 \left( \phi_1(\Delta_1) - \phi_1(\Delta_2) \right) \right).
    \end{equation}
    In our parameter regime, the IQSS protocol has similar or lower performance than the OQSS protocol, and thus in the remaining of this section we focus on the OQSS protocol only.

   In the SWIPHT protocol, there is a notion of a ``harmful" and ``target" transition.
    The difference between these two transitions is that we select the ``harmful" transition to be the transition that we want to drive to obtain a trivial phase.
    The ``target" transition then corresponds to the transition that looks like the target portion of a controlled unitary operation.
    So here we see that there is some freedom in defining which of the two blocks ($1$ or $2$) involves the target and which the harmful transition.
    We define the following choice of sign:
    \begin{align}
        \lambda
        = &  \begin{cases}
                 +1 &  \text{Target is block 1}  \\
                 -1 &  \text{Target is block 2}  \\
        \end{cases}.
    \end{align}
    We also define $\Delta_t$ and $\Delta_h$ based on this choice.
    Then noting that $\omega_p = \Delta_h + \omega_h = \Delta_t + \omega_t$ and defining $\delta\omega_O = \omega_t -     \omega_h$, we can use the definitions for $\theta$, $\phi_1$ as well as trigonometric identities to find $\delta\omega_O/\sigma = \cot\left( \phi_1(\Delta_h)/2 \right) - \cot\left( \phi_1(\Delta_t)/2     \right)$.
    
    Now, if we specify an angle for the two-qubit gate, all of the restrictions up until now allow us to find a pulse frequency and bandwidth that perform two different rotations on each block, but together they combine to form a {\sc cphase} operation.
    We define $\theta_1 = \phi_{00} - \phi_{01}$, $\theta_2 = \phi_{10} -     \phi_{11}$ as well as $\theta = \theta_1 + \theta_2$, $\delta\theta = \theta_1 - \theta_2$.
    These can be written in terms of the harmful/target detunings and $\lambda$ as follows: $\theta = \lambda \left( \phi_1(\Delta_t) - \phi_1(\Delta_h) \right)$ and $\delta\theta = \phi_1(\Delta_t) + \phi_1(\Delta_h)$.
    With these expressions we can then find an expression for the bandwidth in terms of the desired angle
    \begin{align}\label{eq:bandwidth_theta_OQSS_non_resonant}
        \sigma   = &  \lambda \delta\omega_O \frac{\cos(\theta/2)-\cos(\delta\theta/2)}{2\sin(\theta/2)}.
    \end{align}
    
    To successfully generate an arbitrary {\sc cphase} gate we also need to express the control pulse frequency $\omega_p$ in terms of the desired $\theta$ angle.
    In this line, using previous definitions for $\omega_p$, we can write  $\omega_p = \frac{\omega_t + \omega_h}{2} +     \frac{1}{2} \left( \Delta_t + \Delta_h \right)$, which can be easily rewritten in terms of $\theta$ and $\delta\theta$:
    \begin{equation}
        \begin{aligned}
        \omega_p
        = &  \frac{\omega_t + \omega_h}{2} + \frac{1}{2} \left( \Delta_t + \Delta_h \right) \nonumber  \\
        = &  \frac{\omega_t + \omega_h}{2} + \sigma \frac{\sin(\delta\theta/2)}{\cos(\theta/2) - \cos(\delta\theta/2)}.
        \end{aligned}
    \end{equation}
    Using Eq.~\eqref{eq:bandwidth_theta_OQSS_non_resonant} to further simplify the previous equation, we find that the pulse frequency in terms of the desired angle $\theta$ is
    \begin{equation} \label{eqn:offres2pifreq}
    \begin{aligned}
        \omega_p
        = & \frac{\omega_{O,t} + \omega_{O,h}}{2} + \\
        &\lambda \frac{\delta\omega_O}{2\sin\left( \theta / 2 \right)}
        \sqrt{1 - \left( \cos(\theta/2) - 2 \sin\left( \theta / 2 \right) \sigma / \delta\omega_O \right)^2 },
        \end{aligned}
    \end{equation}
    where, in order to ensure that the resulting pulse has finite frequency, we require that the angle of the {\sc cphase} gate is within the range $\theta \in (0,\pi]$.
    Moreover, to make the pulse frequency real, this expression also provides a maximum allowable bandwidth for a given angle $\theta$, $\sigma_\text{max} = \frac{\left| \delta\omega_O \right|}{2} \cot\left( \theta / 4 \right)$.
    
\begin{figure}
        \centering
        \includegraphics[width=\columnwidth]{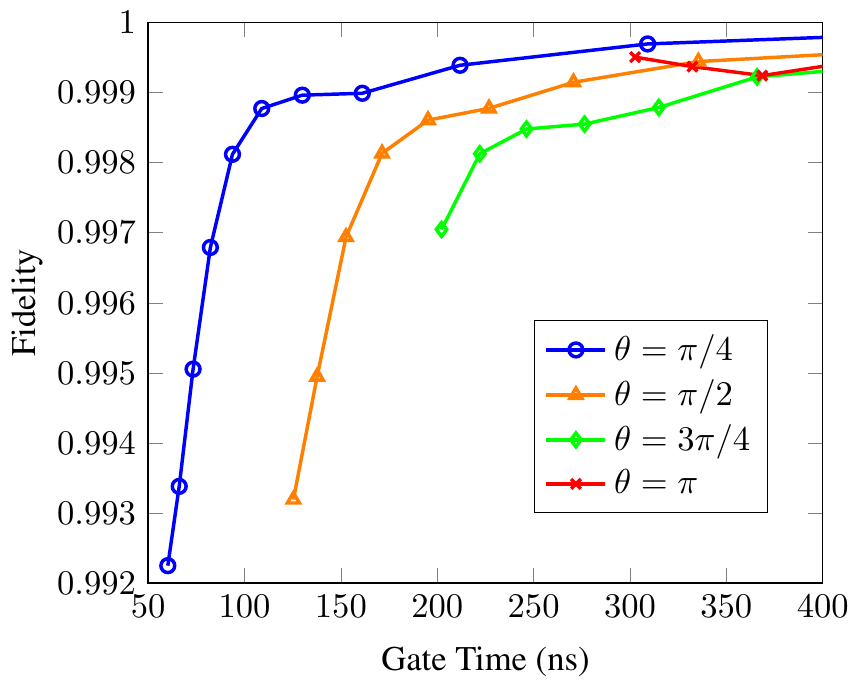}
        \caption{Fidelity of numerical simulations for the {\sc cphase} gate with the OQSS protocol.
        For every angle, there is a maximum allowable bandwidth and hence a minimum allowable gate time.}
        \label{fig:cphase-ad-2pi}
    \end{figure}

    Using this protocol, we find fidelities in excess of $0.999$ and gate times as low as $~60\text{ns}$.
    The numerical evaluations of the fidelity in the simulation for this protocol are shown in Fig.~\ref{fig:cphase-ad-2pi}.
    From the figure we see that the fidelity is consistently above $0.992$ for all angles and gate times.
    By choosing smaller bandwidths, one is able to increase the fidelity.
    The infidelity at small gate times is due to leakage outside the qubit subspace.

    \subsection{{\sc cphase} gate via resonant pulses}\label{subsec:gate-via-resonant}
    
     The construction of the protocols presented in this section is similar to that of the off-resonant protocols in Section~\ref{subsec:gate-via-off-resonant-pulse-oqss}, with the difference that now we require the pulses to be resonant with one of the two transitions in the system.
        In this family of protocols, we will first restrict the value $a = \Omega/\sigma \in \mathbb{Z}$ so that each subspace sees a rotation about the $Z$ axis, independent of     the detuning with the frequency of each transition.
        
        For the IQSS case, this means that the evolution operator is diagonal and has the form $U = \text{diag}\left( f_1 , f_1^* , f_2, f_2^* \right)$ where $f_{j} = {_2}F_1\left(-a,    +a,\frac{-i \Delta_j + \sigma}{2\sigma}, 1\right)$.
        Because we are constructing resonant protocols, without loss of generality, we let the pulse be resonant with the first transition so that $\Delta_1 =0$ and $\Delta_2 =     \delta\omega_I$.
        We may then compute the local invariants, yielding
        \begin{equation}
            \mathbf{G}(U) = \left( \frac{\text{Re}(\alpha)^2}{\left| \alpha \right|^2} , 0 , 2 + \frac{1}{2}\left( \frac{\alpha}{\alpha^*} + \frac{\alpha^*}{\alpha} \right) \right)
        \end{equation}
        where $\alpha = {_2}F_1\left(-a,+a, \frac{-i \delta\omega_I + \sigma}{2\sigma}, 1\right)$ is the Gaussian hypergeometric function.
        Then, demanding that the local invariants of the {\sc cphase} gate (Eq.~\eqref{eq:local_inv_cphase}) be equal to the evolution operator for the pulsed system yields a single equation for this     protocol:
        \begin{equation}\label{eq:root_eq_IQSS_resonance}
            \cos\theta = \frac{ \alpha^2  + \alpha^{*2} }{ 2\left| \alpha \right|^2 }.
        \end{equation}
        As an example, for a $2\pi$ pulse we let $a=1$.
        Then this equation reduces to
        \begin{equation}
            \frac{\delta\omega_I^4 - 6 \delta\omega_I^2 \sigma^2 + \sigma^4}{(\delta\omega_I^2 + \sigma^2)^2} = \cos\theta.
        \end{equation}
        Solving this equation for the bandwidth yields four solutions, and because the bandwidth must be positive, only two are physical based on the sign of $\delta\omega_I$.
        For the $a=1$ case, the final bandwidths are
        \begin{align}
        \sigma_1 = & \left| \delta\omega_I \right| \cot(\theta/4) \\
        \sigma_2 = & \left| \delta\omega_I \right| \tan(\theta/4).
    \end{align}
    This procedure may be repeated for pulses with $a \in \mathbb{N}$, though the resulting equations are more complicated and may be treated numerically.
    Additionally, one may repeat this protocol while using the OQSS transitions.
    The setup is essentially the same, except the OQSS evolution operator will be
    \begin{equation}
        U = \text{diag}\left( 1 , f_1 , 1, f_2 \right).
    \end{equation}
    In this case, the analogous single equation for these protocols will be
    \begin{equation}\label{eq:root_eq_OQSS_resonance}
        \cos\theta = (-1)^a \frac{ 1 + \alpha^2 }{ 2 \alpha },
    \end{equation}
    where now $\alpha = {_2}F_1\left(-a,+a, \frac{+i \delta\omega_O + \sigma}{2\sigma}, 1\right)$
       For OQSS transitions, where the control sech pulse is resonant with a transition partially out of the qubit space, the derivation of the protocols rely on solving Eq.~\eqref{eq:root_eq_OQSS_resonance} for the bandwidth of the pulse.
        In particular, for a sech $2\pi$-pulse ($a=1$) we find that the associated bandwidth that produces a {\sc cphase} gate for a given angle is $\sigma = \left| \delta\omega_O \right| \cot\left( \theta / 2 \right)$,     where again we require that $\theta \in (0,\pi]$.
        On the other hand, if we instead use a $4\pi$-pulse, the solution to the evolution operator has different properties compared to the resonant $2\pi$ case, and we find     that the bandwidth for a specific angle $\theta \in (0,\pi)$ is given by
        \begin{align}
            \sigma
            = \left| \delta\omega_O \right| \frac{\tan(\theta/2)}{ \sqrt{4 + 3 \tan(\theta/2)^2} \pm 2  }.
        \end{align}
        In this case, the choice of sign is arbitrary and the bandwidth does not depend on which transition is designated as the harmful or target.
        However, the choice of sign determines the range of the bandwidth.
        We find that if the sign choice is positive, then $0 < \frac{\sigma}{|\delta\omega_O|} < 1 / \sqrt{3}$ and if the choice of sign is negative, then $1/\sqrt{3} <     \frac{\sigma}{|\delta\omega_O|}$.
        In some protocols derived here, there are multiple ranges of allowed bandwidths.
        These ranges result from the fact that multiple bandwidths satisfy Eq.~(\ref{eq:root_eq_OQSS_resonance}) for a given angle and $\delta\omega$.
        In our parameter regime, this protocol has comparable or lower performance from the others simulated here, so we do not show numerical results in this case.

\begin{figure}
            \centering
            \includegraphics[width=\columnwidth]{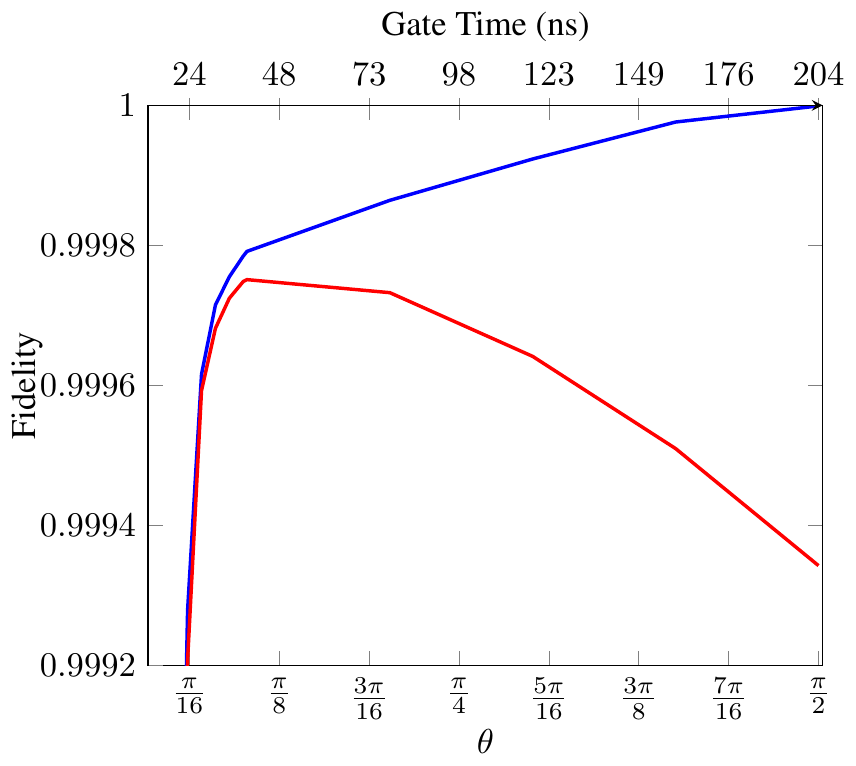}
            \caption{Performance of {\sc cphase} gate from simulation using the resonant IQSS $2\pi$ protocol.
            The upper curve corresponds to using subspace $2$ as the target, $\omega_{I, 2}=\omega_t$, and the bottom curve corresponds to using subspace $1$ as the target,     $\omega_{I,1} = \omega_t$.}
            \label{fig:cphase-2pi-res-iqss}
        \end{figure}

        If we repeat this procedure but now choosing transitions corresponding to the IQSS case, the different bandwidths are obtained by solving Eq.~\eqref{eq:root_eq_IQSS_resonance}.
        For example, in the case of a   $2\pi$-pulse ($a=1$), the bandwidth for this {\sc cphase} gate of angle $\theta \in (0,\pi]$ is $\sigma = \left| \delta\omega_I \right| \cot(\theta/4)$.
        Here we find gates with fidelities as high as $~0.999999$ and gate times as low as     $~24\text{ns}$ for angles in the range of $\pi/16$ to $\pi/2$.
        To construct this protocol, the pulse is driven on resonance with one of the transitions inside the qubit subspace.
        We evaluate the performance of this protocol in simulation by calculating the gate fidelity, shown in Fig.~\ref{fig:cphase-2pi-res-iqss}.
        The two curves correspond to the two different choices of resonant transitions.
        The upper (blue) curve corresponds to the lower right block being the target, and the bottom (red) curve corresponds to the upper left block being the target.
        We find that the fidelity using subspace $2$ as the target is above $0.9998$ for angles from $\pi/8$ to $\pi/2$, and using the other transition as the target produces     lower fidelities of $\sim 0.9995$.
        In either case, we find reasonable gate times for this range of angles.
        The infidelity at smaller angles is due to leakage as a result of larger pulse amplitudes.
        In contrast with the other numerical results, in this protocol the desired angle of the gate fixes the bandwidth and hence the gate time.

        When we repeat this procedure for a $4\pi$-pulse, the {\sc cphase} gate of angle $\theta \in (0,\pi)$, has bandwidth $\sigma = \left|\delta\omega_I\right| \frac{     \tan(\theta/4)     }{ \sqrt{ 4 + 3 \tan(\theta/4)^2 } \pm 2 }$.
        Again, the pulse is driven on resonance with one of the transitions inside the qubit subspace.
        We also find that the range on the bandwidth in the case when the choice of sign is positive becomes $0 < \frac{\sigma}{|\delta\omega_I|} < \frac{1}{2+\sqrt{7}}$ and when     the choice of sign is negative, we have $\frac{1}{\sqrt{7}-2} < \frac{\sigma}{|\delta\omega_I|}$.

\begin{figure}
            \centering
            \includegraphics[width=\columnwidth]{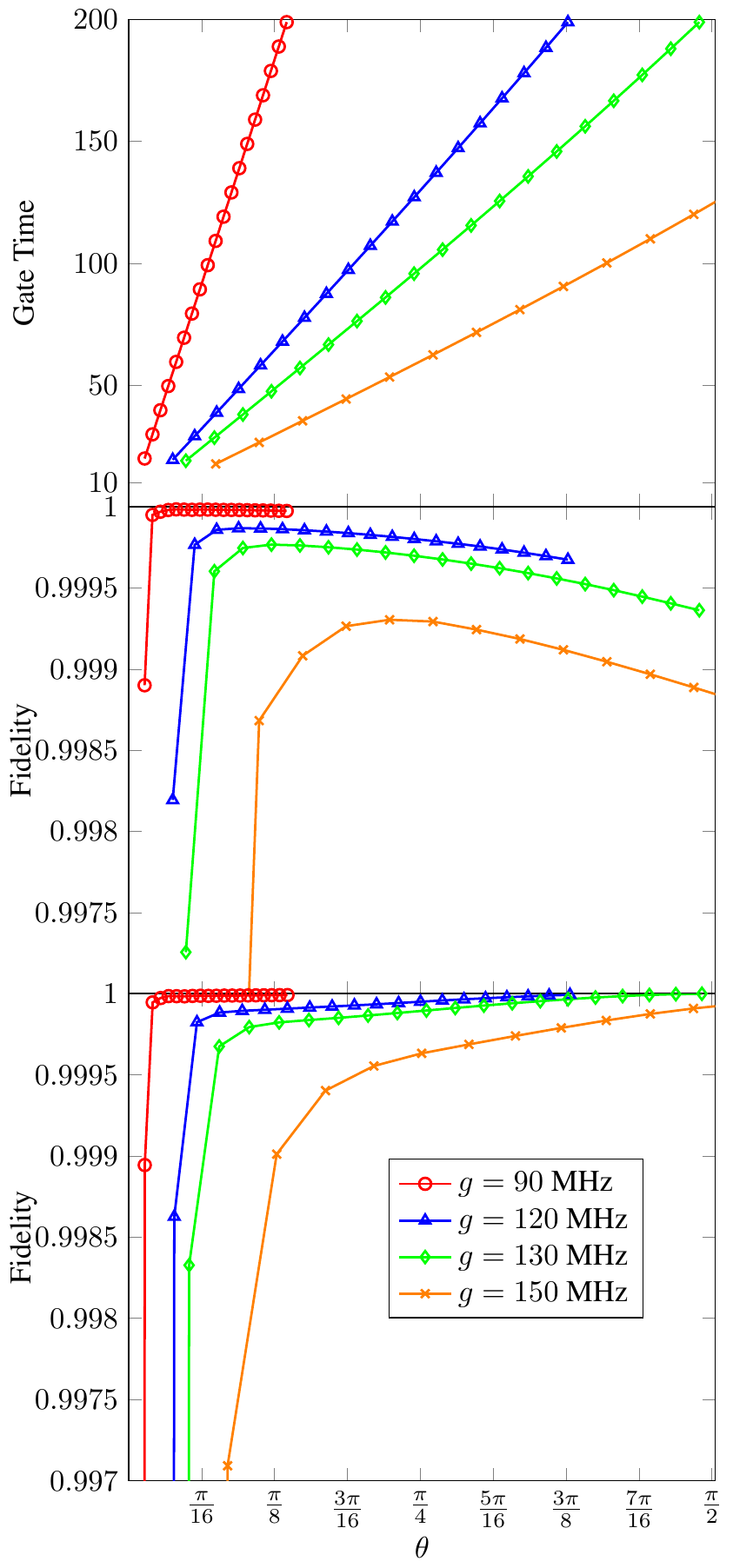}
            \caption{Properties and performance of the {\sc cphase} gates using the IQSS $2\pi$ protocol over a range of coupling strengths.
            The upper panel shows the gate time for each coupling strength $g$ as a function of the angle of the {\sc cphase} gate.
            The middle panel shows the fidelity at each angle for the various coupling strengths using subspace $1$ as the target, $\omega_{I,1} = \omega_t$.
            The lower panel shows the fidelity at each angle for the various coupling strengths using subspace $2$ as the target,     $\omega_{I,2} = \omega_t$.
            For all panels, gate times between $~10\text{ ns}$ and $~200 \text{ ns}$ are considered.}
            \label{fig:coupling}
        \end{figure}

        So far we have fixed the coupling to $g = 130 \text{ MHz}$.
        Now we focus on the IQSS $2\pi$ protocol and evaluate its performance as a function of the coupling strength.
        We determine two primary features as we vary the coupling strength.
        Firstly, weakly coupled systems produce gate times that increase rapidly as a function of the desired angle, as shown in Fig.~\ref{fig:coupling}.
        Secondly, increasing the coupling strength decreases the fidelity, as shown in Fig.~\ref{fig:coupling}.
        Overall, we find that for a range of coupling strengths we are able to find high fidelities exceeding $0.998$.
        In some cases the fidelity is as high as $0.999999$.
        In all cases, the fidelity drops for smaller angles due to leakage as a result of larger pulse amplitudes.
        We limit these simulations to $200\text{ ns}$ gate durations to compare the different coupling strengths because this protocol has no upper bound on the gate time.

    \subsection{{\sc cphase} protocols comparison}\label{subsec:protocols-comparison}

    In Table~\ref{table:cphasesummary} we provide a summary of the results of the various protocols.
    Overall, we find that there is flexibility in the way of constructing {\sc cphase} gates.
    For instance, one does not necessarily need to drive on resonance with one of the transitions.
    Additionally, one may choose various pulse areas or bandwidths for different implementations.
    For instance, one may choose to derive protocols with $a = 3, 4, \dots$.
    This has the potential to reduce the duration of the gate at the cost of potentially introducing more leakage, which one could possibly address by incorporating DRAG~\cite{motzoi_simple_2009, gambetta_analytic_2011, motzoi_improving_2013} into the pulse design, but this is beyond the scope of this work.
    In terms of performance, the two best protocols are the OQSS arbitrary frequency via $2\pi$-pulse and IQSS resonant $2\pi$-pulse protocols.
    Comparing the fidelities and gate times for a range of angles, if a small angle is desired one should choose the IQSS resonant $2\pi$-pulse protocol because at small angles it provides consistently higher fidelities ($\sim 0.9999$ compared to $\sim 0.998$) at comparable gate times, and sometimes fidelities as high as $0.999999$.
    On the other hand, if a larger angle is desired, the OQSS arbitrary frequency via $2\pi$-pulse protocol is preferable due to its flexibility in the bandwidth, yielding potentially lower gate times ($\sim 120 \text{ ns}$ compared to $\sim 200 \text{ ns}$).
    The other protocols produce fidelities on the order of $\sim 0.98$ generally due to their higher bandwidths, which result in more leakage.
    In systems that do not have higher available states, these protocols may be more useful as they can produce smaller gate times for a range of angles.
    
    \begin{table*}[t]
        \centering
        \caption{Summary of results for various {\sc cphase} protocols.
        These results are derived in Sections~\ref{subsec:gate-via-off-resonant-pulse-oqss}-\ref{subsec:gate-via-resonant}.
        Each row denotes the analytical results of a particular protocol producing a {\sc cphase} gate by an angle $\theta$.
        The way to read this column is from left to right.
        First choose a pair of transitions corresponding to the transitions in Figure~\ref{fig:elevels}, and then choose a pulse area.
        The properties of the selected pulse will then be the bandwidth for a particular angle and the range on such bandwidths.
        In some cases, there are two disconnected ranges of allowable bandwidths.
        The OQSS $2\pi$ protocol has an arbitrary bandwidth in the sense that it is not a function of the desired angle.
        However, the bandwidth must satisfy the constraint in the corresponding ``Bandwidth Range" column.}
        \vspace*{+3mm}
        \begin{tabular}{ l | c | c | c | c }
            Transitions &  \text{Pulse Area}       &  \text{Frequency}              &  \text{Bandwidth}                    &  \text{Bandwidth Range}   \\
            \hline
            IQSS & $2\pi$                    &  $\omega_t$ &  $\sigma = |\delta\omega_I| \cot(\theta/4)$       & $0 < \sigma < + \infty$   \\
            \hline
            IQSS & $4\pi$                    &  $\omega_t$ &  $\sigma_\pm = |\delta\omega_I| \frac{\tan(\theta/4)}{\sqrt{4 + 3 \tan(\theta/4)^2} \pm 2 }$       & $0 <     \frac{\sigma_+}{|\delta\omega_|} < \frac{1}{2+\sqrt{7}} < \frac{1}{-2+\sqrt{7}} < \frac{\sigma_-}{|\delta\omega_I|} < +\infty$   \\
            \hline
            OQSS & $2\pi$                    &  $\omega_t$ &  $\sigma = |\delta\omega_O| \cot(\theta/2)$       &  $0 < \sigma < + \infty$   \\
            \hline
            OQSS & $4\pi$                    &  $\omega_t$ &  $\sigma_\pm = |\delta\omega_O| \frac{\tan(\theta/2)}{\sqrt{4 + 3 \tan(\theta/2)^2} \pm 2 }$       & $0 <     \frac{\sigma_+}{|\delta\omega_O|} < \frac{1}{\sqrt{3}} < \frac{\sigma_-}{|\delta\omega_O|} < +\infty$   \\
            \hline
            OQSS & $2\pi$                    &  \text{Eq. }~\ref{eqn:offres2pifreq}&   \text{Arbitrary}                  & $0 < \sigma < \frac{|\delta\omega_O|}{2} \cot(\theta/4) < +\infty$
        \end{tabular}
        \label{table:cphasesummary}
    \end{table*}

    \section{Single Qubit Gates}\label{sec:sq-gates}
    Now we turn our attention to single qubit operations.
    Before we proceed, there is one thing to note about systems with an always on interaction such as ours.
    In this section we develop single qubit operations that are performed in the presence of another qubit.
    As we saw before, the interaction between the two qubits dresses the energy eigenstates, and the Hamiltonian in the dressed basis obtains an effective $Z\otimes Z$ coupling.
    In this way, single qubit gates are manifestly less well defined than in the case where the effective $Z\otimes Z$ coupling is much smaller (or not present).
    In contrast, this distinction is less important in systems with a smaller always-on interaction because the dressed and bare bases are closer together.
    For the development of our single qubit operations, we choose to work in the dressed basis because it is closer to what would actually be used in an experimental setting.

    We develop a set of arbitrary single qubit rotations of the form $R_{\hat{n}} (\theta) = e^{-i \theta \hat{n} \cdot     \vec\sigma /2 }$, which can be generated by combining $R_{\hat{x}} (\theta)$ and $R_{\hat{z}} (\pi/2)$ rotations.
    Since rotations about the $Z$ axis may be produced by shifts in the frequency of the microwave pulse~\cite{mckay_efficient_2017} with zero gate time and no loss in fidelity, we only consider the development of the rotations about the $x$-axis.
    Without loss of generality, we can write the desired evolution operator for such a rotation as $\mathbb {I} \otimes R_{\hat{x}} (\theta)$.
    To develop these gates, we consider sequences of square pulses so that the Hamiltonian is simply the Hamiltonian in Eq.~\ref{eq:hqss} for piecewise constant $\Omega_i(t)$ and $\omega_{p,i}(t)    $.
    The evolution operator for the qubit subspace can be written as
    \begin{align}
        U(\vec{\tau},\vec{E},\vec{\omega}) = &  \begin{pmatrix}
                                                    U_1(\vec{\tau},\vec{E},\vec{\omega}) & 0 \\ 0 & U_2(\vec{\tau},\vec{E},\vec{\omega}),
        \end{pmatrix}
    \end{align}
    where $U_j(\vec{\tau},\vec{E},\vec{\omega}) = \prod_{i=1}^{N} U_{j,i}(\tau_i,E_i,\omega_{p,i})$ and $U_{j,i}(\tau_i,E_i,\omega_{p,i})$ is the evolution operator for the     $j^\text{th}$ block over the duration of the $i^\text{th}$ square pulse.
    The $i^\text{th}$ square pulse has duration $\tau_i$, pulse amplitude $E_i$ and frequency $\omega_{p, i}$.

    Instead of solving exactly for parameters of each pulse that perform the desired evolution on each subspace, we define an objective function to optimize which is $f_{\hat{n},    \theta} (\vec{\tau},\vec{E},\vec{\omega}) = F\left( \mathbb{I}\otimes R_{\hat{n}}(\theta) , U(\vec{\tau},\vec{E},\vec{\omega}) \right)$ where $F(U,V)$ is the fidelity between two unitary operators.
    We do this for several reasons.
    Primarily, there is no guarantee that such solutions exist, and even if they did, they would likely not be simple.
    Moreover, even if we solve for a sequence of pulses that exactly implements the desired evolution, in simulation and experiment the fidelity will not be exactly $1$     due to decoherence.
    In practice we use global, constrained optimization algorithms over the $3N$ parameters to find such sequences of pulses.
    The region in which the optimization is performed is determined by experimental limitations such as ramp-up times for the microwave pulses on the order of $~1\text{ ns}$ and maximum possible amplitudes of each pulse based on the microwave pulse generators of about $~20\text{ MHz}$.

    The desired evolution operator for the qubit subspace here is $\mathbb{I} \otimes R_{\hat{x}}(\theta)$ so that $U_j(\vec{\tau},\vec{E},\vec{\omega}) = R_{\hat{x}}(\theta)$ for each $j=1,2$.
    Without loss of generality, we choose a sequence of pulses resonant with the first subspace so that $\omega_{p,i} = \omega_{I,1}$.
    Then, with $E_i \in     \mathbb{R}$, this sequence of square pulses naturally produces rotations about the $x$-axis for the first subspace, $R_{\hat{x}}(\theta) = U_1(\vec{\tau},\vec{E},\vec{\omega})    $.
    This provides the constraint $\theta / 2 = d_1 \sum_{i=1}^{N} \tau_i \left| E_i \right| \label{eq:sqXcons}$.
    Now the optimization is over $2N$ parameters with one constraint.

\begin{figure}
        \centering
        \includegraphics[width=\columnwidth]{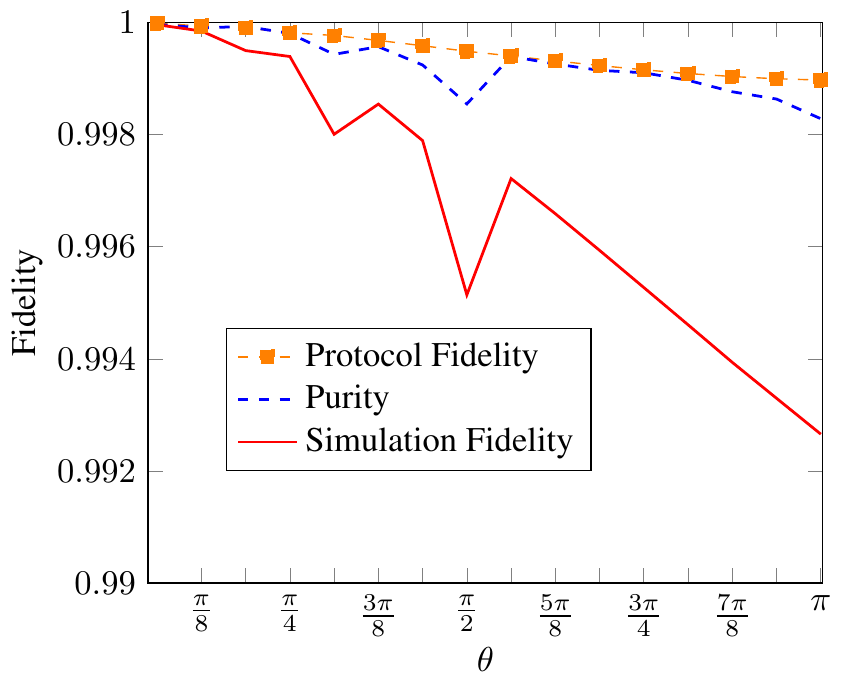}
        \caption{Performance of single qubit rotation protocol.
        The upper (orange) curve is the Protocol Fidelity, determined by global optimization over the parameters of a sequence of resonant square pulses for the $4$-level system.
        The middle (blue) curve is the purity as a function of the angle.
        The bottom (red) curve is the fidelity found from simulation of the full time dynamics of the system after using the result from the protocol as an initial condition in a local optimization.}
        \label{fig:sqXRot}
    \end{figure}

    We evaluate the performance of the single qubit $X$ rotation protocol.
    This involves two steps: The first step is to determine the parameters on some sequence of square pulses by the optimization of $f_{\hat{n},\theta} (\vec{\tau},\vec{E},\vec{\omega})$, which yields what we define as the ``Protocol Fidelity", see Fig.~\ref{fig:sqXRot}.
    The second step is to take the resulting sequence of square pulses and simulate the full time dynamics of the system, using a local optimization to improve the results of the protocol in the simulation.
    This is done by using the parameters of each pulse sequence from the protocol as initial conditions to a local optimization algorithm that improves the fidelity.
    We refer to this as the ``Simulation Fidelity" in Fig.~\ref{fig:sqXRot}.
    We find ``Simulation Fidelities" above $0.992$ for all angles $0 < \theta \leq     \pi$.
    All of these gates have durations from $\sim 15\text{ ns}$ to $\sim 25 \text{ ns}$.
    Because there is a gap between the ``Simulation Fidelity" and purity in the figure, we see that there is some coherent error occurring.
    This is due to coupling to higher excited states which are not included in the $4$-level system and is the primary cause of the infidelity.
    The dip at $\theta = \pi/2$ is due to the fact that we use a local optimizer for the ``Simulation Fidelity" and the curve is not guaranteed to be smooth.

    \section{Conclusions}\label{sec:conclusions}

    Using the analytical evolution operator for the hyperbolic secant pulse acting on a two-level system, we have derived a collection of {\sc cphase} gates for transmon qubits.
    We have demonstrated that these gates produce high fidelities typically in excess of $\sim 0.999$ and in some cases as high as $0.999999$ and typical gate times less than         $\sim 100\text{ ns}$.
    Moreover, we show that one of these protocols is robust in the fidelity for a range of angles and coupling strengths $g$.
    Finally, we demonstrate that arbitrary single qubit gates may be achieved via microwave pulses in this realistic parameter regime using sequences of square pulses.
    In conclusion, we produce high-fidelity parameterized entangling gates that may be applied in realistic systems for use in quantum simulation algorithms.

    \section{Acknowledgements}\label{sec:ack}
    This research was supported by the Department of Energy, Award No. de-sc0019318 and by the Army Research Office, Award No. ARO W911NF1810114.

    \begin{appendix}
        \section{Hyperbolic Secant Pulse Solution}\label{sec:app-sech}
        The basis for the {\sc cphase} gate is the analytic solution for the evolution operator of a 2-level system driven by a hyperbolic secant pulse.
        The pulse is defined as     $\Omega(t) = \Omega_0 \sech(\sigma t)$, where $\sigma$ is the pulse bandwidth and $\Omega_0$ is the pulse strength.
        One can show that the form of the Hamiltonian for a given transition in the interaction frame is
        \begin{align}
            H_2(t)
            = &  \begin{pmatrix}
                     0 &  \Omega(t) e^{i\Delta t} \\
                     \Omega(t)^* e^{-i\Delta t} & 0
            \end{pmatrix},
        \end{align}
        where $\Delta$ is the detuning of the pulse with the $\ket{0}\leftrightarrow\ket{1}$ transition, i.e. $\Delta = \omega_p - \epsilon_1$.
        By following a previous discussion of this problem~\cite{economou_proposal_2006}, we define $c = \frac{1}{2}\left( 1+ i \frac{\Delta}{\sigma} \right)$, $a = \frac{\Omega}{\sigma}$, $\zeta = \frac{1}{2} \left( 1 + \tanh(\sigma t) \right)$, $\alpha(a,c,\zeta) =\ {_2}F_1 (a,-a,c^*,\zeta)$, and $\beta(a,c,\zeta)  =\ {_2}F_1 (a+1-c,1-a-c,2-c,\zeta)$ where ${_2}F_1$ is one of Gauss' hypergeometric functions.
        Then the evolution operator is
        \begin{align}
            U(t,-\infty)
            = &  \begin{pmatrix}
                     \alpha(a,c,\zeta) & -\frac{i a}{c} \zeta^c \beta(a,c,\zeta)   \\
                     -\frac{i a}{c^*} \zeta^{c^*} \beta(a,c,\zeta)^* &   \alpha(a,c,\zeta)^*
            \end{pmatrix}.
        \end{align}
        Here the initial condition is $U(-\infty,-\infty) = \mathbb{I}$, though in practice we take the initial time to be some finite value that is sufficiently large for our results to converge.
        Since we are only interested in the end result of the pulse, we consider the evolution operator at $t=+\infty$,
        \begin{align}
            U & = U(+\infty,-\infty) \nonumber   \\
            & = \begin{pmatrix}
                    {_2}F_1\left(-a,a;\frac{\sigma -i \Delta }{2 \sigma };1\right)            & -i \text{sech}\left(\frac{\pi  \Delta }{2 \sigma }\right) \sin (a \pi ) \\
                    -i \text{sech}\left(\frac{\pi  \Delta }{2 \sigma }\right) \sin (a \pi ) & {_2}F_1\left(-a,a;\frac{i \Delta +\sigma }{2 \sigma };1\right)
            \end{pmatrix}.
        \end{align}
        Then it is clear that for $a \in \mathbb{Z}$, the evolution operator is diagonal.
        In this instance, we can express the evolution operator with $U = \text{diag}\left\{     e^{-i\phi_a} , e^{+i\phi_a} \right\}$.
        If we use a $2\pi$-pulse (i.e. $a=1$), $\phi_1(\Delta) = 2 \arctan\left(\frac{\sigma}{\Delta}\right)$.
        If instead we consider a     $4\pi$-pulse, then $a=2$ and $\phi_2(\Delta) = 2 \arctan\left( \frac{4\Delta/\sigma}{(\Delta/\sigma)^2-3} \right)$.

        \section{Local invariants}\label{sec:mi}
        To develop our {\sc cphase} gates, we will use the local invariants~\cite{makhlin_0000,Zhang2003} of unitary operations in $SU(4)$, denoted here as $G_i$ for $i \in \{1,2,3\}$.
        These are three quantities that may be computed from any element of $SU(4)$, and are invariant under operations in $SU(2)$.
        That is to say that for $\mathcal{U}\in SU(4)$ and $V_i \in SU(2)$,
        \begin{equation}
            G_i \left( V_1 \otimes V_2 \mathcal{U} V_3 \otimes V_4 \right) = G_i(\mathcal{U}).
        \end{equation}
        Therefore, the local invariants convey the nonlocal properties of the operator $\mathcal{U}$ and give a unique representation of any class of two-qubit gates that are equivalent up to local operations.
        Note that single qubit rotations about the $Z$ axis may be efficiently performed for transmons~\cite{mckay_efficient_2017}, and that hyperbolic secant pulses can produce no population transfer, as discussed in Appendix~\ref{sec:app-sech} and in Ref.~\onlinecite{economou_high-fidelity_2012}.
        These two facts allow us to develop protocols for {\sc cphase} operations that only consider their nonlocal characteristics and have no overhead in terms of the fidelity or time required to perform the single qubit gates associated with their local-equivalence classes.
        Hence, the local invariants of the analytical unitary evolution, $U$, for our four-level system driven by a hyperbolic secant pulse will be the starting place for constructing our protocols for a {\sc cphase} gate.
        The quantities $G_i$ are obtained by first placing $U$ in the magic basis~\cite{Bennett1996} defined by the unitary transformation~\cite{makhlin_0000,Zhang2003} 
        \begin{equation}
            Q=\frac{1}{\sqrt{2}}\left(\begin{array}{cccc}{1} & {0} & {0} & {i} \\ {0} & {i} & {1} & {0} \\ {0} & {i} & {-1} & {0} \\ {1} & {0} & {0} & {-i}\end{array}\right).
        \end{equation}
        The local invariants are the coefficients of the characteristic polynomial of the matrix $M(U) = (Q^\dagger U Q)^T (Q^\dagger U Q)$, and they are given by the following expressions:
        \begin{equation}\label{eq:local_invariants}
        \begin{aligned}
            G_1 = & \text{Re} \frac{\tr\left( M(U) \right)^2}{16 \det U}, \\
            G_2 = & \text{Im} \frac{\tr\left( M(U) \right)^2}{16 \det U}, \\
            G_3 = & \frac{ \tr\left( M(U) \right)^2 - \tr\left( M^2(U) \right) }{4 \det U}.
            \end{aligned}
        \end{equation}

    \end{appendix}

    \bibliography{references-manual}

\end{document}